\documentclass[onecolumn]{aa}
\usepackage{graphicx}
\usepackage{epsfig}
\begin{document}
\title{High accuracy proper motions in crowded fields}
\author{C. Alard}
\offprints{C. Alard}
 \institute{Institut d'Astrophysique de Paris, 98bis boulevard Arago, 75014 Paris France, 
\email{alard@iap.fr}
\and
  Observatoire de Paris, 77 avenue Denfert Rochereau, 75014 Paris France.}

\date{}

\abstract{  
The image subtraction method is a powerful tool to analyze 
the light variations in crowded fields. This method is
able to achieve a nearly optimal differential photometry, even
in very dense regions. However, image subtraction is not limited 
to photometry, and it is shown in this paper that the method
can be generalized to the measurement of the differential
proper motions. It is important to emphasize that image 
 subtraction can re-construct an un-biased frame to frame astrometric 
transform. A nearly optimal determination of this astrometric transform 
can be performed by expanding the spatial variations of the kernel using 
the derivatives of the constant kernel solution. It is demonstrated that 
an expansion using first and second derivatives is optimal.
Differential refraction can be corrected easily using an artificial  
image and the first derivatives of the kernel. To illustrate the 
ability of the method to measure proper motions, a small sub-area 
near the center of a Galactic Bulge field covered 265 times
by the OGLE II experiment has been selected. The resulting astrometric 
accuracy is very close to the photon noise for the faint objects, while 
for the brighter ones it approaches closely the limit set by the residual  
seeing fluctuations on the smaller scales. The accuracy obtained with this 
new method is compared to the accuracy achieved using classical methods. 
The improvement obtained varies from a factor of $\simeq$ 2 for the brighter  
objects, to more than a factor of 3 for the fainter ones. This improvement 
is typical of the improvement that was obtained for the measurement of the 
photometric variations.
   }
   \maketitle

\section{Introduction}
The development of the image subtraction method was motivated by the need
to perform an optimal analysis of the light variations in crowded
stellar fields. In particular, the large amount of data produced by the 
microlensing surveys towards the Galactic Bulge or the LMC (Udalski {\it et al.}
1994 (OGLE), 1997 (OGLE II), Alcock  {\it et al.} 1997 MACHO, Alard \& Guibert 1997 (DUO))
posed the problem of the optimal detection and measurements of the light
variations of a faint object surrounded by a dense background of stars.
The first successful attempts were performed by Tomaney \& Crotts (1996) 
who developed a Fourier based image subtraction method. This work was followed by a least-square implementation of the image subtraction method
 Alard \& Lupton (1998). This least-square implementation is nearly optimal 
provided that the kernel solution is almost constant across the image. However
this is not always the case, and sometimes it can be a severe limitation.  
As a consequence the least-square method was generalized to the case of 
spatially variable kernel solution with conservation of flux (Alard 2000). 
The ability of the least-square method to obtain nearly optimal results
was demonstrated at many occasions (see for instance Alard 1999, or Wozniak
2002). It was also demonstrated that this method improved the photometry
by a factor 2 to a factor 4, with respect to classical
methods (see Alard 1999 for details). It was recently suggested by Paczy\'nski
(2002), that it might be possible to extend this improvement to the measurement
of the stellar differential motions. As shown by Eyer \& Wozniak (2001) it is possible
 to detect and estimate the proper motions of stars using the version of image
subtraction optimized for photometric measurements. However, it is clear that
to perform a nearly optimal astrometric measurement, it necessary to extend
and develop substantially  the image subtraction method.
\section{Basic equations}
An observed image is the result of the application of several physical process
to the ``true image''. The three major process are instrumental and atmospheric
blurring, the sampling introduced by the detector array, and the noise. Each
 of these process can be described appropriately by simple mathematical
operations. It is of particular importance to image subtraction that
the instrumental and atmospheric blurring of the image can be described 
appropriately be a convolution process. As a result it should be possible
to transform an image to exactly the same observing condition as another
one by a convolution process. The basic idea of image subtraction
is that this convolution kernel can be written as sum of linear functions.
Considering that the kernel $\phi$ takes the values $\phi_{i,j}$ 
on the image 2 dimensional grid, and that the basis functions $f^k$ takes
the corresponding values $f^k_{ij}$ on the same grid, one can write:
$$
 \phi_{ij} = a_{ij} f^k_{ij}
$$
 in order to simplify the notation, the summation symbol will be omitted. 
 The following implicit summation rule will be used: {\bf summation occurs when the
index is repeated twice.}
\\\\
In general the coefficient $a_j$ of the kernel decomposition will depend upon
the position in the image, thus:
\begin{equation}
  \phi_{ij} = a_j(x,y) \ f^k_{ij}
\end{equation}
By applying this kernel transform to the best seeing image $G$, it is possible
to match closely the other image. Assuming Gaussian statistics the optimal
kernel transform should minimize the chi-square between the convolved image
and the bad seeing image $B$.
$$
 \chi^2=\left|\left|\frac{a_k G_{i-l,j-m} f^k_{lm} - B_{ij}}{\sigma_{ij}}\right|\right|^2
$$
 To simplify the notations it is interesting to define the convolved 
 basis functions:
$$
 W_{ij}^k = G_{i-l,j-m} f^k_{lm}
$$
Thus:
\begin{equation}
 \chi^2=\left|\left|\frac{a_k W_{ij}^k - B_{ij}}{\sigma_{ij}}\right|\right|^2
\end{equation}
\section{Image subtraction and optimal astrometry}
\subsection{modeling of differential motions}
To perform optimal astrometry one has to measure the displacements of
stars between 2 images with the best possible accuracy. The displacements
due to proper motions of stars corresponds to tiny shifts on the image
grid. It is possible to model such small shifts in a given image
 $I$ with great accuracy
using a first order expansion in spatial coordinates:
\begin{equation}
 \delta I_{ij} = \left({\frac {\partial I}{\partial x}}\right)_{ij} dx +  \left({\frac {\partial I}{\partial y}}\right)_{ij} dy
\end{equation}
 Thus to model the displacements between our 2 images it is necessary to 
 introduce the local derivatives. Our model of the image is (see eq 2):
$$
 M_{ij}=a_k W_{ij}^k
$$
According to eq (3) the shifted model $\tilde M$ can be written:
$$
 \tilde M_{ij} = M_{ij} + \left({\frac {\partial M}{\partial x}}\right)_{ij} dx +  \left({\frac {\partial M}{\partial y}}\right)_{ij} dy
$$
 Note that:
$$
\left({\frac {\partial M}{\partial x}}\right)_{ij} =   {\frac {\partial }{\partial x}} \left( B_{i-l,j-m} \ \phi_{lm} \right) \simeq {\frac {\partial }{\partial x}} \int B(u,v) \phi(x-u,y-v) \ dudv \simeq \int B(u,v) \ {\frac {\partial }{\partial x}} \ \phi(x-u,y-v) \ dudv
$$
Consequently:
$$
 \left({\frac {\partial M}{\partial x}}\right)_{ij} \simeq B_{i-l,j-m} \left({\frac {\partial \phi}{\partial x}} \right)_{lm} 
$$
Thus, we see that it is possible to model small shifts between the image
by including the derivatives of the kernel in the kernel expansion. Provided
 that the kernel basis is complete this basis should also contain the
kernel derivatives (as a linear combination of the original basis function).
As a consequence it is clear that the general image subtraction method should 
be able to model small shifts between the images. Since the kernel decomposition
depends on the position in the image, the shift may also depend on the
position. However, we may wonder if this mapping of the shifts between
the images corresponds to the right astrometric registration.
 \subsection{Image subtraction and astrometric alignment}
 One condition for the perfect astrometric alignment between 2 systems
 is that there should be no systematic offset between the 2 systems.
 Thus it is important to check that the subtracted image does not contains
 some systematics astrometric shift. A residual shift in the subtracted
 image would mean that there is a displacement between the convolved
 solution and the image to fit. Assuming that the unshifted subtracted image
 is $R_{ij}$ , and that the bad seeing image $B_{ij}$
 is shifted by $(\delta_x,\delta_y)$ with respect to the convolved 
 good seeing image $a_k W_{ij}^k$, using eq (3) one can write that\ the shifted subtracted $S_{ij}$ image will be: 
$$
 S_{ij}=a_k W_{ij}^k - B_{ij} = R_{ij}+\left(\frac{\partial B}{\partial x}\right)_{ij} \delta x +  \left(\frac {\partial B}{\partial y}\right)_{ij} \delta y
$$
Thus the estimation of the displacement  is once
again a linear least-squares problem related to the estimation of 
$(\delta x, \delta y)$. Here the 2 least-squares vectors are
the derivatives of the image. The corresponding  normal least square equation is a linear system of equations with the following form:
\begin{equation}
 N \delta_{xy} = V
\end{equation}
Where $N$ is a matrix containing the cross products of the vectors.
V is a vector which components are the cross products of the vectors
with the subtracted image.
For the ideal, unshifted solution, $R_{ij}$, $\delta_{xy}=0$, and consequently
 , according to eq (4), the cross products: 
$$
\frac{R_{ij}}{\sigma_{ij}^2} \left(\frac{\partial B}{\partial x} \right)_{ij} {\rm and} \ \ \frac{R_{ij}}{\sigma_{ij}^2} \left(\frac{\partial B}{\partial y} \right)_{ij}
$$
have to be zero. It is possible to demonstrate the same
result for the shifted subtracted image. The image subtraction method
implies that  $\chi^2$ must be minimal with respect to any the
kernel coefficient $a_p$, using eq (2) one can write:
\begin{equation}
 \frac {\partial \chi^2}{\partial a_p} = 0 = W_{ij}^p \left[\frac{a_k W_{ij}^k - B_{ij}}{\sigma_{ij}}\right] = W_{ij}^p \ R_{ij}
\end{equation}
As it was already shown in Sec. 3.1 the kernel basis of functions should
contain the kernel derivatives. A convolution with the kernel derivatives
will recreate the derivatives of the image. Thus the basis of functions
 $W_{ij}^p$ should contain the image derivatives. Then using eq (5) it is
 easy to prove that the scalar product of the image derivatives with the
 subtracted image is zero. And consequently using eq (4), we see that
the displacement $\delta_{xy}$ has to be zero.
Thus there is no systematic bias on in an astrometric
registration performed using the image subtraction method.
\section{Optimal astrometric alignment}
The general image subtraction method is able to perform an unbiased 
image alignment, however the least-square process, and in particular
 the spatial variability involves a large number of free parameters which
may introduce unnecessary noise. To perform an optimal astrometric alignment
it is required to use the smallest possible set of parameters. 
 Image subtraction with spatial variations requires many parameters.
To estimate the number of parameters, it is necessary to multiply the numbers
of parameters for the constant kernel solution by the number of polynomial
coefficients used the spatial expansion. Typically, a constant kernel
solution requires about 50 parameters and with a spatial expansion of
second degree we get 300 parameters. However, if we consider a smaller
region of the image, we expect that the spatial variation of the kernel
will be small, and that we might be able to model the kernel variations
using an expansion in Taylor series of the constant kernel solution 
 $\phi^0$ ($\phi^0$ can be for instance the solution near the center
of the frame:
\begin{equation}
 \phi_{ij} = \phi^0_{ij}+\sum_{l <= k} a_{kl}(x,y) \left [\frac {\partial^{k+l} \phi^0}{\partial x_k \partial y_l} \right]_{ij} 
\end{equation}
It is particularly interesting to use a Taylor series of degree 2. In this
expansion, the first derivatives will take into account the motion between
the frames, while the second derivatives will help to reduce the general
chi-square. It is important 
to note that the components corresponding to the astrometric displacement
$(\frac{\partial \phi^0}{\partial x}, \frac{\partial \phi^0}{\partial y})$ are orthogonal to the other components of the kernel expansion. For
instance:
$$
\int_{-\infty}^{\infty} \frac{\partial \phi^0}{\partial x} \phi^0 dx = \left[\frac{1}{2} \left(\phi^0 \right)^2\right]_{-\infty}^{\infty} = 0 {\rm \ \ \ \  (Since \ the \ kernel \ decreases \ exponentially)}
$$
Similarly, one can demonstrate easily that the first derivatives are
orthogonal to the second derivatives. This property is also true for
the convolved basis functions. The convolution with the kernel derivatives
will give the derivatives of the convolved image. Thus, our former reasoning
will apply. Except, that here the relevant integration boundaries will be
the edges of the image. The image may not be zero on its edges, but what
matter for the value of the integral is the difference between the edges.
Statistically if the value on the edges are un-correlated, the mean
should be zero. In any case the cross-product of the first derivative
with the 2 others component should be very small. Thus, the least-squares
vectors corresponding to the kernel derivatives are orthogonal to the
other least-square vectors. In the least-square normal equations, the
matrix contains the cross products of the vectors, and thus many components
of the matrix will be zero. This matrix will be non-zero only in local
blocks. In particular, the block corresponding to the first derivatives will
be isolated, and surrounded with zeros. As a result the least-square solution
for the first derivatives will be independent from the estimation of the
other parameters. This property is very interesting, because it means that
the errors will be also independent on the errors on the other parameters, and
as a consequence should be minimized. Note that formally the scalar
product of the vectors, represent the least-square matrix element in
case the statistically weight $\sigma_{ij}$ is constant (gaussian noise).
In case of Poissonian weighting, the difference should be negligible in
most area, except for bright stars. But even around bright stars, provided
the psf is nearly symmetrical, one can show that in this case the integral
is close to zero.
Thus it means that the errors on estimation of the astrometric displacement
will be minimal. This interesting property of the spatial dependence in
terms of derivatives could be completely flawed if the spatial expansion 
did not conserved the flux. Hopefully it is easy to demonstrate that
the derivatives have zero sums. For instance:
$$
 \int_{-\infty}^{\infty} \frac{\partial \phi}{\partial x} dx =\left[ \phi \right]_{-\infty}^{\infty} = 0 {\rm \ \ \ \ Since \ the \ kernel \ decrases \ exponentially}
$$
The same reasoning apply to all the other derivatives. Thus, using eq (5) we
see immediately that the integral of the kernel as a function of position
is constant, and is equal to the integral of the constant kernel solution.
As a conclusion, we see that the Taylor expansion of order 2 provides a natural
 basis to expand the spatial variations of the kernel: the errors 
 on the differential motions are minimized, and flux conservation
 is preserved. 
\section{Correction of differential refraction}
 To first order, the effect of differential refraction is to shift
slightly the position of the stars. The amplitude of the shift depends 
on the color of the stars. Thus to correct the differential refraction,
one must know the color of the objects in the image. To compute the
color of the objects it is necessary to run a source extraction code
in the V and I band. The magnitudes in each band are estimated by PSF
fitting. Once the color and magnitude of each star are known, it is
possible to correct for the differential refraction effect. For instance
assuming a simple model where the displacement is proportional to the
color, the correction will be proportional to the color of the star
multiplied by its flux. In this simple model, to compute
the differential refraction correction it is sufficient to reconstruct
an artificial image where the amplitude of the stars are multiplied by
their color, and convolve it with the kernel derivatives. A least-square
fit of these 2 vectors will give directly the 2 coefficients corresponding
to the differential refraction along x and y. Obviously the method can
be extended to non-linear modeling of the color versus differential refraction
effect.
\section{Application to OGLE data}
\subsection{practical implementation}
 The OGLE II project (Udalski  {\it et al} 1997) produced a large amount of good quality 
 CCD image in either Galactic or Magellanic fields. Most of the data is 
 in the I band, with also some images in V and B. 
 Some of the Galactic Bulge fields offers an impressive
coverage of the observing seasons. In particular the field OGLE Bulge SC 33 
has been observed 265 times over an interval of approximately 1200 days. This
data set is very well suited to experiment the new method presented in this
paper. In this field the density of stars is very large, and thus it is
sufficient to study a smaller sub-area of the field. The area selected
is near the center of the SC 33 field, the Galactic coordinates of
the center of the field are: (l$=$2.2, b$=$-3.7). An image surface
of 512$\times$512 pixels which corresponds to about 
$3.5^{'}\times 3.5^{'}$ on the
sky is large enough to give a total of about 800 stars for which the
color and magnitudes can be measured with decent accuracy. Although
it is important to note that due to the severe blending, this may
not be true for the fainter objects, which actually can be unresolved 
blends. Due to this source confusion issue, it is interesting
to have the best possible spatial resolution in the reference image.
One of the image was taken under outstanding seeing conditions, with
low airmass. This image has also the advantage to be near the
middle of the data time interval, which minimize the differential motion
of the stars. All the frame were aligned
to this reference frame (see Alard \& Lupton 1998, and Alard 2000
for details), and then subtracted. The constant
kernel solution was derived by a local fit on the bright stars.
The solution that was selected is the local solution 
 having the best $\chi^2$ for the whole image.
 The derivatives of the constant kernel solution were
expanded spatially using polynomials of 2nd degree. The differential
refraction was corrected using a polynomial of the same degree. Thus,
the total number of components to fit on the image using linear least-square
is 43. In order to minimize the errors due to the drift scan procedure,
 the frame was splited into 3 sub-frames along the y direction
. Using the full set of 265 subtracted images, it is possible to 
produce astrometric curves along each direction for the 782 stars of the
 sample. Knowing the baseline flux of the star $F_0$, the local
kernel solution $\phi$, and the local psf on the reference image,
$\psi$, one can relate the displacement to the residuals in the subtracted
image $S_{ij}$ using eq (3):
$$
  S_{ij}= F_0 \left({\frac {\partial \eta}{\partial x}}\right)_{ij} dx +   F_0 \left({\frac {\partial \eta}{\partial y}}\right)_{ij} dy \ \ \ \ {\rm \ with: \ \ \ \ } \eta_{ij} = \phi_{i-j,k-l} \psi_{k,l}
$$
Thus dx and dy can be measured by a 2 parameter linear least-square fit: 
\begin{equation}
 S_{ij}= a_0 \left({\frac {\partial \eta}{\partial x}}\right)_{ij} + a_1 \left({\frac {\partial \eta}{\partial y}}\right)_{ij} \ \ \ \ {\rm \ with: \ \ \ \ } dx=\frac{a_0}{F_0}, \ dy=\frac{a_1}{F_0}
\end{equation}
\begin{figure}
 \epsfig{file=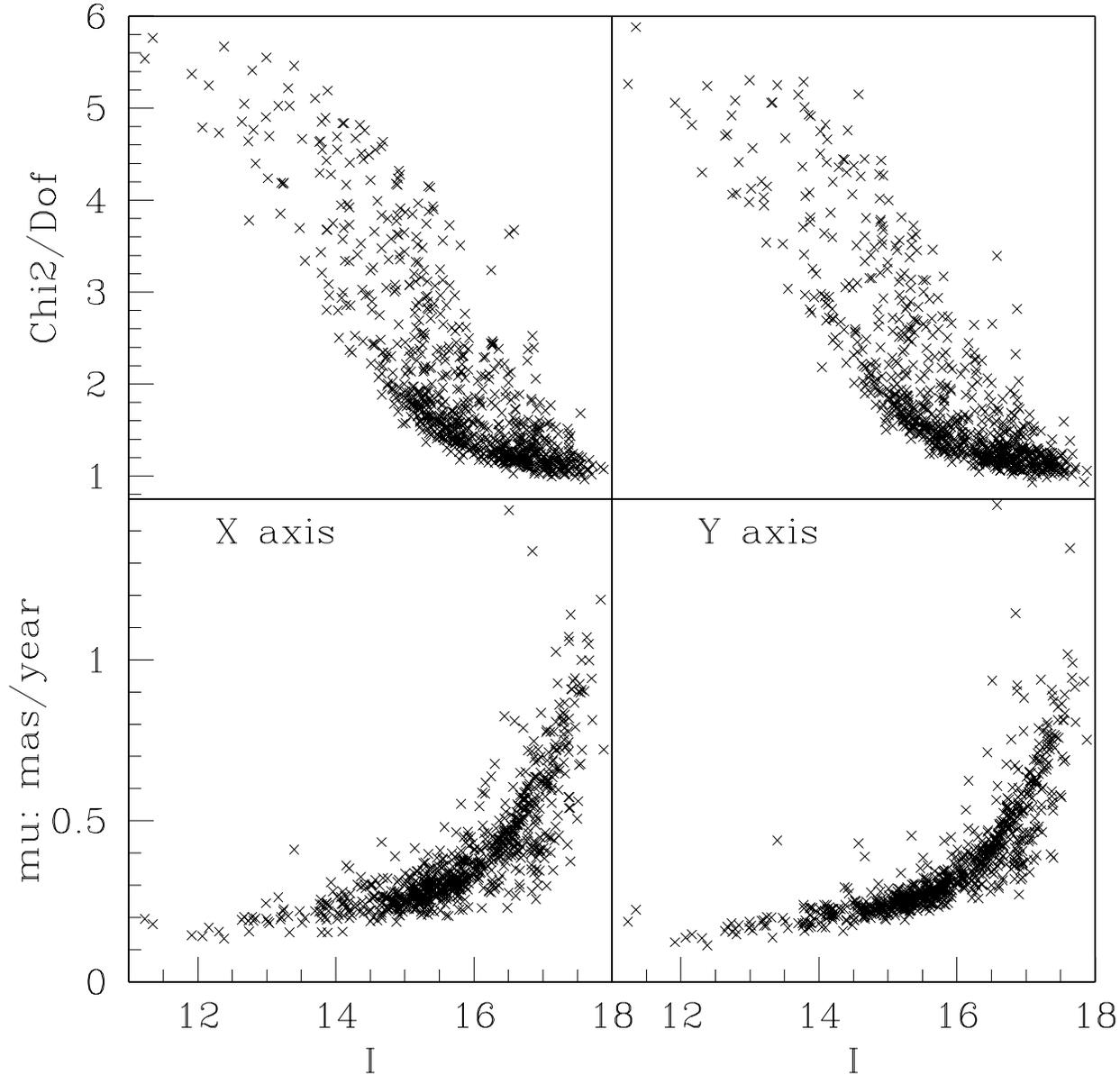,width=17cm}
 \caption{The error on measuring the proper motion along the X and Y directions as a function of the I band magnitude for the 782 stars of the sample.  The
diagrams in the bottom presents the real value of the errors, while on
the top the errors are compared to the theoretical poissonian
expectations.}
\end{figure}
\begin{figure}
 \epsfig{file=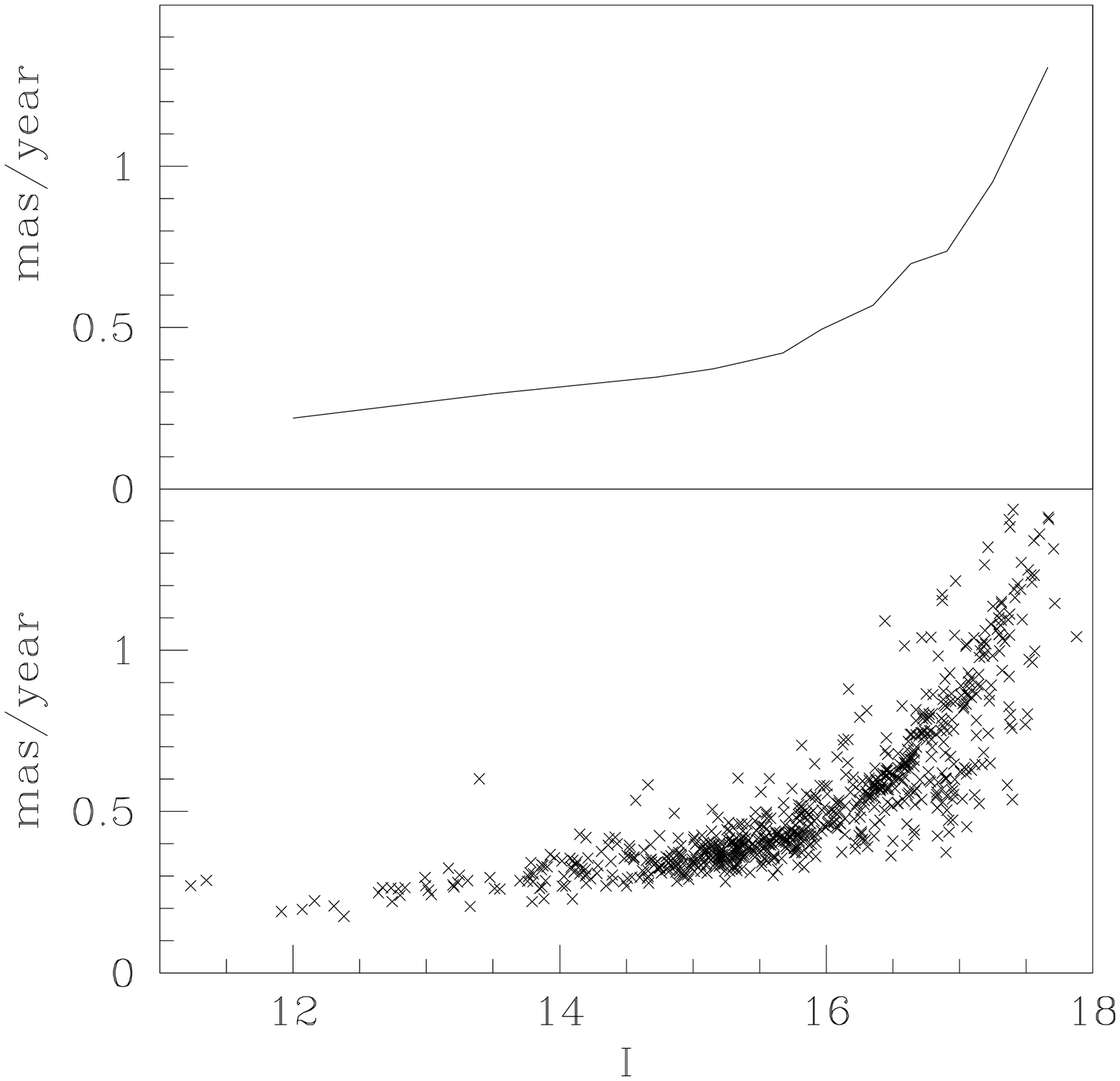,width=14cm}
 \caption{The error on measuring the proper motion as a function
of the I band magnitude for the 782 stars of the sample. Note that
one star near the magnitude I$=$13 has a large deviation from the rest
of the sample. Although this star is unsaturated, it is very close 
to a very bright saturated star. The diagram at the bottom presents
the errors for each stars, while the diagram on the top gives
the mean error.}
\end{figure}
 The differential motions of each star in each of the 265 frames was estimated
 using eq (6). To estimate the proper motion of the star, the curve representing
 its displacement as a function of time was fitted with a straight line. This fit
 gives also the opportunity to compute the error on the estimation of the slope
 (proper motion). The brightness of the star $F_0$ was derived by PSF fitting
 (see also Sec. 5).
\subsection{The errors on the proper motion}
 In principle the error on the slope should be derived directly
 from the Poisson fluctuations. However,  the real errors exceeds the Poisson
 expectations, and especially for bright stars (see Fig. 1). Obviously, there
 are other causes of uncertainties that are larger than the Poisson fluctuations
 for bright stars. These cause do certainly include the seeing fluctuations,
 and possibly also, the variation of sensitivity on the
 surface of a pixel, the noise in the flat fielding, and finally the drift
 . It is simple to analyze each of these causes separately.
 First the seeing fluctuations: in Alard \& Lupton (1998) it was shown that
 on a scale of an arc minute the variation of the centroid position due to the
 seeing fluctuations was about 10 mas. However, since in the present work
 it is possible to correct the systematic spatial variations between the
 2 images to subtract, at least, the large scale seeing fluctuations 
 should be corrected. Although, it is clear that nothing can be done
 on smaller scales: typically a few arc seconds. To be more specific
 the seeing fluctuations cannot be corrected on a distance smaller
 than the minimum separation between 2 resolved stars, which is about
 10 pixels $\simeq$ 4 arc seconds. Since the seeing
 fluctuations between 2 stars separated by $\theta$ decreases like $\theta^{1/3}$
 , and that there are 265 data points, a quick estimate of the error gives 
 $\simeq (4/60)^(1/3)*10/\sqrt{265} \simeq$ 0.25 mas. The other errors except
 possibly the uncertainties in the flat fielding are much smaller. Numerical
 simulation shows that the variations of sensitivity on the surface of a pixel
 should be at least one hundred time smaller than the errors due to the
 seeing fluctuations. In case the drift scan errors would contribute significantly
 to the error budget, we would see a different between the errors in each direction.
 But Fig. 1 shows that this is not the case. The last possibility to investigate
 are systematic flat fielding errors on the scale of the PSF. For a  1 arc second
 seeing, a systematic flat fielding variation (with the right shape) of 1 \% 
 would translate into
 a variation of 7 mas of the centroid, corresponding to a global error
 on the proper motion of about 0.4 mas. But it is important to emphasize,
 that it is unlikely to find systematic variation in the flats
 on the scale of the PSF, and that it is less likely again that these
 systematic variations have the right shape (the shape of the derivatives).
 As a conclusion it is very likely that most of the excess in the astrometric
 error for bright stars is due to residual seeing fluctuations on smaller scales.
\subsection{Testing the accuracy of the method}
The recent release by Sumi {\it et al.} (2003) of a large scale study
of the proper motions of stars in a large number of OGLE Bulge fields
offers the possibility to compare
the accuracy achieved using this method to the accuracy obtained using
 classical methods (DoPHOT).
By comparing Fig. 2 in this paper to Fig. 4
 in Sumi {\it et al.} (2003), we can estimate that for clump giants (I $\simeq$ 15.5)
 Sumi {\it et al.} obtain a mean accuracy slightly above 1 mas  $\simeq$ 1.2 mas, while
 Fig. 2 in this work gives an accuracy of $\simeq$ 0.4 mas. This improvement of factor
 of 3 in the accuracy is also typical of the improvement obtained for the photometry.
 Note that the brighter unsaturated stars this improvement is closer to a factor of
 2, while for the fainter stars it is a little above 3. One may wonder why such an
 improvement is observed even on the bright stars. The answer might be that in 
 classical methods to perform an optimal measurement of the position of a star
 it is neccessary to build a psf model for each new image. Image subtraction
 does not need a psf model. The measurement of the motions in the subtracted 
 images, requires to build a psf model only for the reference image. There
 is no need to re-create a psf model for each new image. As the calculation of
 the position is quite sensitive to the psf model, this result certainly in
 a reduction of the noise.
 As a final testing it is
 interesting to investigate the structure of the proper motions as a function of color
 and magnitudes. The proper motions along the Galactic longitude and latitude (l,b) have
 been represented in Fig. 3 as a function of color. The bluer stars belong to the
 disk population, and show a systematic motion a along the l axis as expected from
 the differential rotation between the the Galactic disk and bulge. No systematic 
 motion are observed along the b axis.
 Fig. 3 includes all stars in the sample, whatever the accuracy achieved on these
 objects, there was no attempt to perform any cut-off in the error distribution. 
 Since the disk stars have systematic positive motions in longitude,
 it should be possible to reject the disk stars and to retain only the bulge
 stars by selecting stars having negative motions in l. Stars having negative
 l motions have been selected by means of a 3 $\sigma$ cuts. Their position
 in the color-magnitude diagram is presented in Fig. 4. Two different cuts
 have been used. The first cut in the lower panel favors the maximum number
 of stars by allowing the fainter star to enter the selection. 
 The second panel presents stars having accurately measured
 proper motions, thus allowing a lower cut-off, but also restricting 
 the selection to the brighter objects. An interesting feature start to appear
 in the lower panel: there are a few stars just above the main sequence. Even
 if the kinematics of these stars overlaps the Bulge kinematics these stars
 are probably not Bulge stars. The upper panels reveals their true nature:
 we see a well defined sequence, just parallel to the main sequence but
 just a little above. These stars are most likely wide binaries. These
 binaries belongs to the disk population, but the long period motion
 of their center of gravity alter their proper-motions, to the point
 that they overlap some of the Bulge kinematics. The ability of the
 method to find these stars indicates that the method is really accurate.
 In the lower panel of Fig. 4 there are also some stars below the main
 sequence. These stars could be white dwarfs. A further examination of
 a larger sample will help to confirm this possibility. As a final test
of these proper motions, it is interesting to compare the velocity dispersion
of the Bulge stars to the recent results obtained by Kuijken \& Jones (2002).
Most of the clump giants in this field belong to the Bulge population. Thus
this is a good set of stars to estimate the Bulge velocity dispersion.
By selecting stars in a small window around the clump it is possible to derive
the following (l,b) proper motion dispersions: (2.76, 2.16) mas. 
Kuijken \& Jones (2002) found (2.9, 2.5) mas for a field located close
to OGLE SC 33. It is difficult to determine whether the smaller dispersion
found in this study is related to the accuracy of the measurement of the
proper motions, or is due to a different sampling of the Bulge stellar
population. Assuming a distance to the Bulge of 8 Kpc, the proper motion
dispersions measured in this study translate to respectively (105, 82) km/s.
This result is consistent with the former study of Spaenhauer, 
Jones \& Whitford (1992).
\begin{figure}
 \epsfig{file=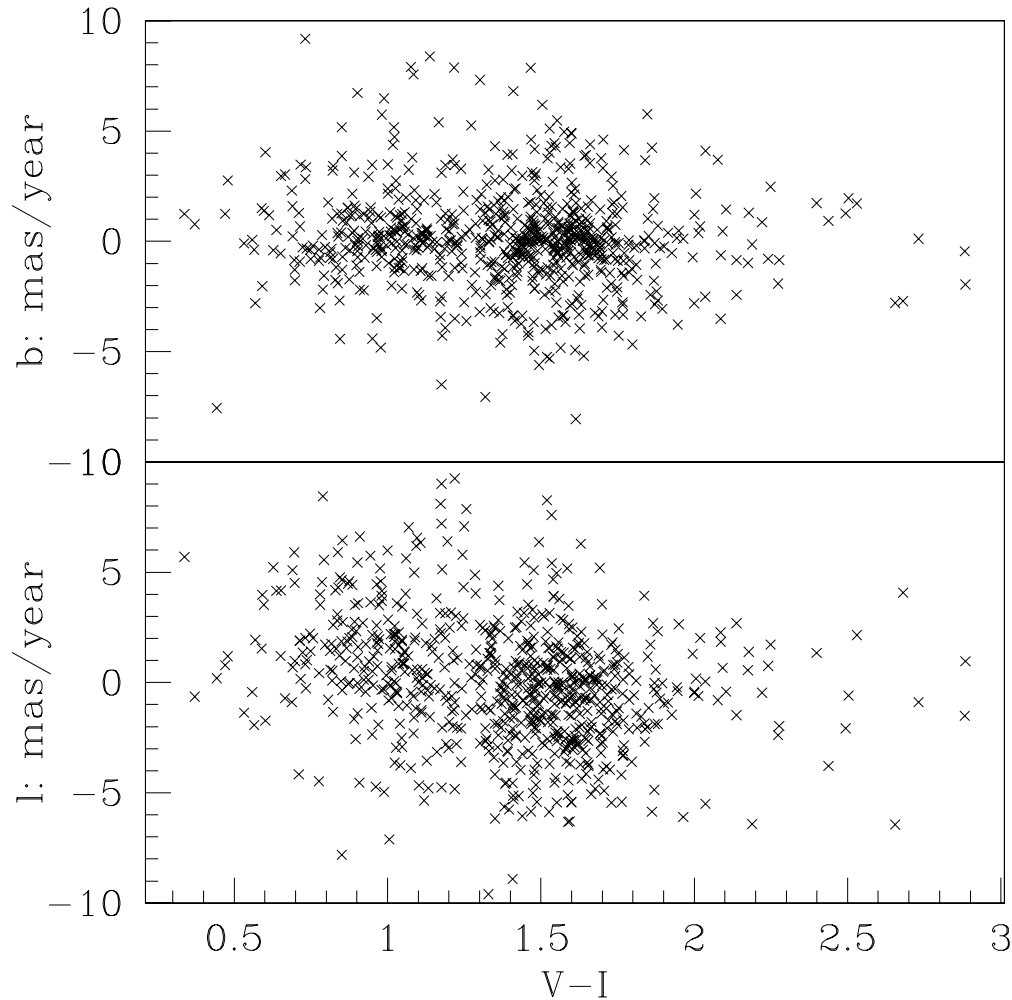,width=14cm}
 \caption{The proper motion of the of all the stars in the sample along
the (l,b) axis as a function of color. Note that the proper motions of the 
bluer stars are systematically shifted along the l axis, but not along
the b axis.} 
\end{figure}
\begin{figure}
 \epsfig{file=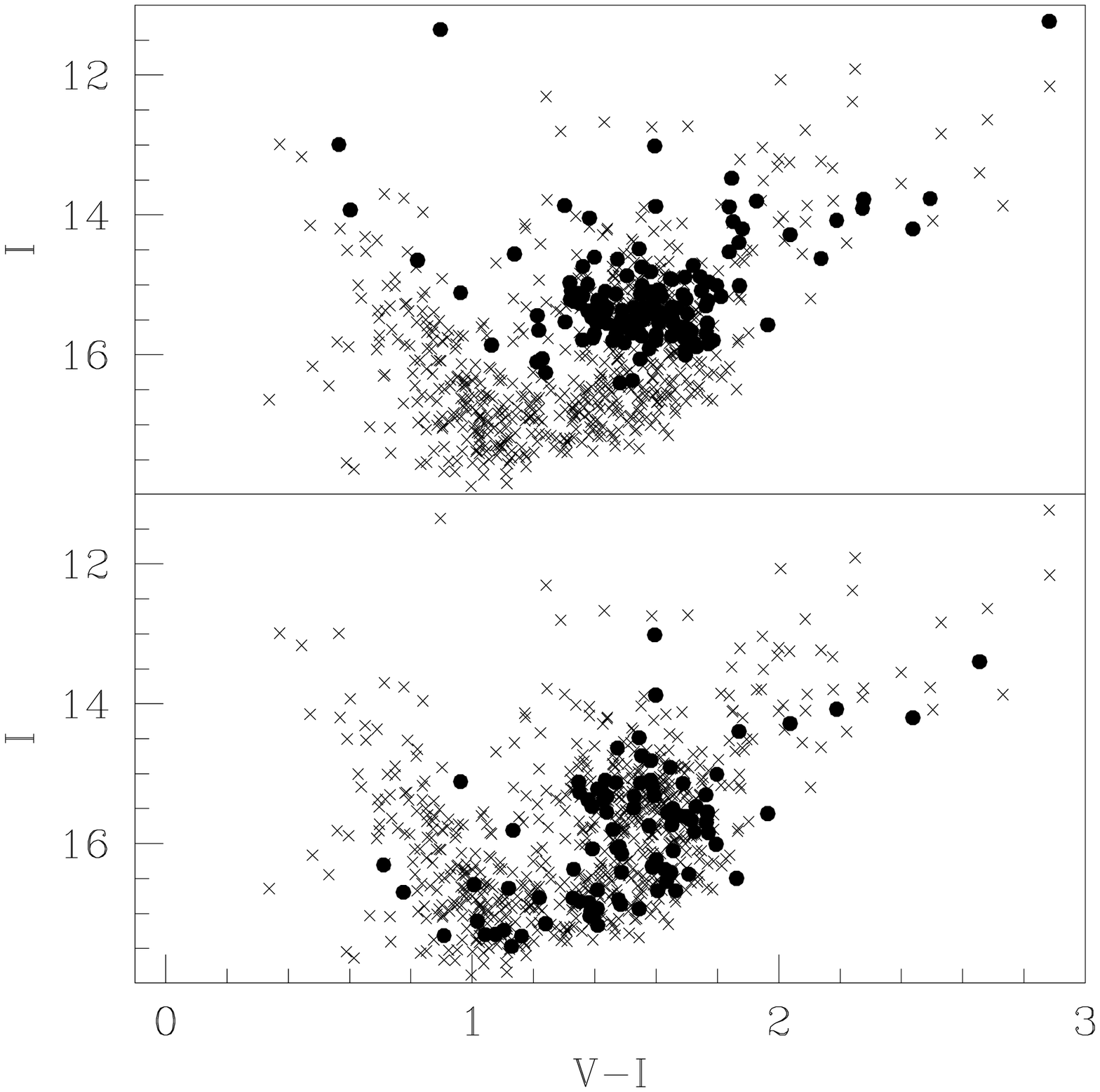,width=17cm}
\caption{Selection of stars having negative proper motions along the l direction.
These stars should belong to the Bulge population. The cross represents all the stars
while the bold dots represents negative l motion.
Two different cuts
 have been performed: $\mu_l < -1$ plus error on $\mu < 0.35 \simeq 3 \sigma$ below 0 (top), and $\mu_l < -2.5$ plus error on $\mu < 0.83$ (bottom). In the lower diagram,
the Bulge sequence is well outlined. The top of the Bulge main sequence, as well
as the Bulge sub-giants and giants are clearly visible. There also some stars
below and above the main sequence. The the star below the main sequence could be
white dwarfs. The star above the main sequence appear more clearly in the upper
diagram (lower cut-off, better accuracy). These stars could be wide binaries.} 
\end{figure}
\acknowledgements{The author would like to thank B. Paczy\`nski for interesting
discussions and suggestions. The stay of the author in Princeton 
was supported by the NSF grant AST-1206213, and the NASA grant
NAG5-12212 and funds for proposal \#09518 provided by NASA through a grant from
the Space Telescope Science Institute, which is operated by the Association of
Universities for Research in Astronomy, Inc., under NASA contract NAS5-26555. 
}
\end{document}